\documentclass[12pt]{article}
\usepackage{latexsym}
\usepackage{graphicx}
\usepackage{times}

\begin{document}

\title{Finding way to bridge the gap between quantum and classical mechanics}
\author{Wang Guowen\\College of Physics, Peking University, Beijing, China}
\date{December 12, 2005}
\maketitle

\begin{abstract}
We have calculated the momentum distributions of nanoparticles in
diffraction and interference dependent on the effective screening
mass parameter or size parameter and presented the calculations
for a nanoparticle inside an infinite square potential well and
for a tunnelling nanoparticle through a square potential barrier.
These results display the transition from quantum to classical
mechanics and the simultaneous wave-particle duality of
nanoparticles. The concept that the effective screening effect
increases with increasing size of an object paves way for
development of nanomechanics and nanotechnology.
\end{abstract}

\section{Introduction}
It is well known that there is a gap between quantum and classical
mechanics. Nanomechanics is required for the development of
nanotechnology that was proposed first in 1959 by Richard Feynman
[1]. Now it becomes possible to find way to bridge the gap due to
the understanding of quantum reality and interference phenomena as
presented in Ref.[2]. In the article a particle is described as a
non-spreading wave packet satisfying a linear equation within the
framework of special relativity and quantum interference
experiments are explained with a hypothesis that there is a
coupling interaction between the peaked and non-peaked pieces of
the wave packet. It has also been mentioned that concerning a
macroscopic object, for example, a tiny grain of sand, roughly
speaking, because the outer matter in it, like a barrier, screens
nearly completely the off-peak part of the inner matter, the
diffraction and interference of the grains fundamentally do not
take place when they pass through slits. The concept that the
effective screening effect increases with increasing size of an
object gives a logical description of transition from quantum to
classical mechanics and paves way for development of nanomechanics
and nanotechnology. This article will present some typical
consequences of the concept to display the transition and the
simultaneous wave-particle duality of nanoparticles. The
explanation of the transition in terms of environment-induced
decoherence proposed by such as Zurek [3] seems to be untenable.

\section{Diffraction and interference of nanoparticles}
A grain of matter in size larger than 100 nm is generally
considered as a macroscopic object and a particle up to 1 nm as a
perfect quantum particle such as fullerene $\texttt{C}_{60}$ [4].
In order to illustrate the behavior of nanoparticles we consider a
spherical nanoparticle as a model as shown in
Fig.1(a).\begin{figure}[htbp]
\centerline{\includegraphics[width=3.2in,height=2.447in]{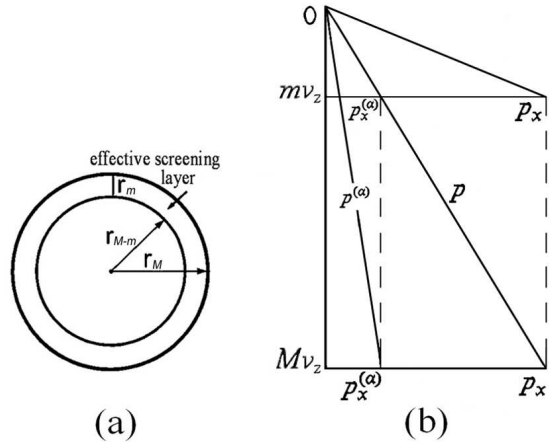}}
\label{Fig.1} \caption{(a) Sketch of the effective screening layer
in a spherical nanoparticle. (b) Illustration of momentum
components of the diffracted nanoparticle.}
\end{figure}
Let \emph{m} denote the mass of the effective screening layer,
\emph{M} the total mass and $\alpha=m/M$ the effective screening
mass parameter (ESMP). Furthermore, let $\rho$ denote the average
mass density of the spherical nanoparticle, $r_M$ the radius of
the nanoparticle and $r_m$ the thickness of the effective
screening layer, so $M=4\pi r_M^3\rho/3$ and $M-m=4\pi
r_{M-m}^3\rho$/3. Therefore the ratio $\sigma=r_m/r_M$ as an
effective screening size parameter (ESSP) is
\begin{equation}
\label{1} \sigma=\frac{r_m}{r_M}=\frac{r_M-r_{M-m}}{r_M}
=\frac{M^{1/3}-(M-m)^{1/3}}{M^{1/3}}=1-(1-\alpha)^{1/3}
\end{equation}
\begin{equation}
\label{2} \alpha=1-(1-\sigma)^3
\end{equation}
Since the thickness of the screening layer having quantum behavior
can be assumed to have a value around a certain value, say 5 nm,
the larger the size of the nanoparticle is, the more it is like a
classical particle. Now we can investigate diffraction of
nanoparticles from a single slit and interference from a double
slit in terms of the screening effect.

\subsection{Diffraction of nanoparticles from a single slit}
Assuming that \emph{x} axis is perpendicular to a slit in the slit
plane and \emph{z} axis perpendicular to the slit plane, according
to quantum mechanics, if the wave function at the slit is
$\psi(x)$, the diffraction of a particle can be calculated by
using the Fourier transformation [5]
\begin{equation}
\label{eq3} \phi(p_x)=\frac{1}{\sqrt{2\pi \hbar}}\int
\exp(-\frac{ip_x x}{\hbar})\psi(x)\texttt{d}x
\end{equation}
The momentum of the diffracted particle is
\begin{equation}
\label{eq4} p=\sqrt{m^2 {v_z}^2+{p_x}^2}
\end{equation}
where $v_z$ is the velocity component of the particle along
\emph{z} axis. For a nanoparticle with the ESMP $\alpha$, its
momentum is
\begin{equation}
\label{eq5} p^{(\alpha)}=\sqrt{M^2v_z^2+p_x^{(\alpha)2}},\ \mbox{
}p_x^{(\alpha)}=\alpha p_x
\end{equation}
where $p_x^{(\alpha)}$ is the diffraction contribution of the
effective screening layer and is equal to $\alpha p_x$ as
illustrated in Fig.1(b).

The normalized function of the single slit of width, say, 50 nm,
as shown in Fig.2(a) is
\begin{equation}
\label{eq6} \Psi(x)=\frac{1}{\sqrt{50}}, \ \mbox{ }|x|<25
\end{equation}
\begin{figure}[htbp]
\centerline{\includegraphics[width=4in,height=1.493in]{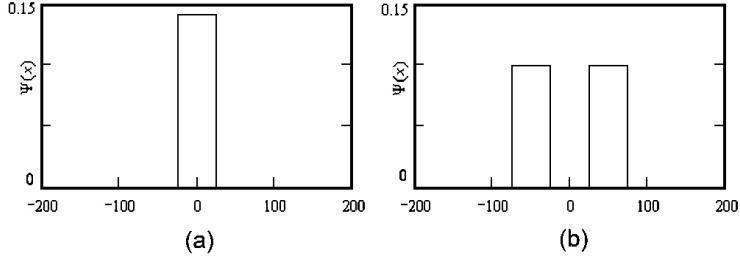}}
\label{Fig.2} \caption{Slit functions: (a) at the single slit, (b)
at the double slit.}
\end{figure}
\begin{equation}\label{eq7}
\Psi(x)=0, \ \mbox{ }|x|>25
\end{equation}
Substituting $p_x^{(\alpha)}/\alpha$ for $p_x$ into Eq.3, we have
the normalized momentum wave function
\begin{equation}
\label{eq8}\Theta(p_x^{(\alpha)})=\frac{1}{\sqrt{2\pi
\hbar}}\int\exp(-\frac{ip^{(\alpha)}_xx}{\alpha\hbar})\Psi(x)\texttt{d}x
\end{equation}
and the normalized momentum distribution of the nanoparticle:
\begin{equation}\label{eq9}
P(p_x^{(\alpha)})=\frac{|\Theta(p_x^{(\alpha)})|^2}{\int
|\Theta(p_x^{(\alpha)})|^2 \texttt{d}p_x^{(\alpha)}}
\end{equation}
Taking $\hbar=1$ in the natural unit, the distributions
$P(p_x^{(\alpha)})$ with different values of $\alpha$ are shown in
Fig.3.
\begin{figure}[htbp]
\centerline{\includegraphics[width=4.5in,height=2.486in]{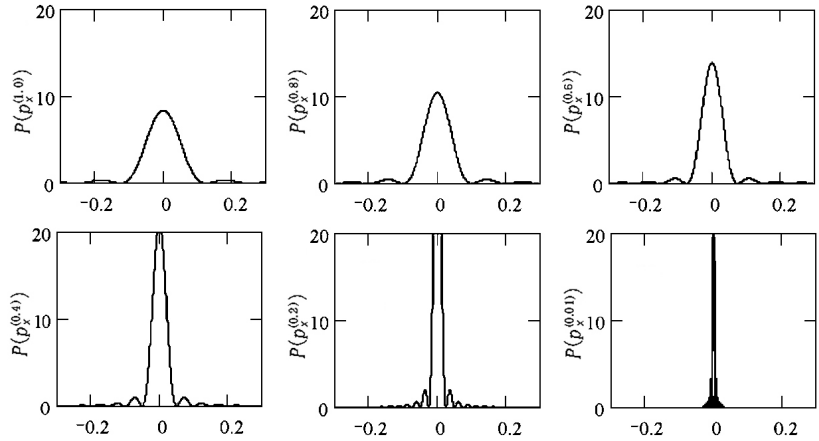}}
\label{Fig.3} \caption{The normalized momentum distributions
$P(p_x^{(\alpha)})$ of the nanoparticle in the single slit
diffraction with different effective screening mass parameters
$\alpha$.}
\end{figure}
We see the width of the distribution decreases to its classical
limit as $\alpha\rightarrow 0$, instead of wrong $\hbar\rightarrow
0$.

\subsection{Interference of nanoparticles from a double slit}
Likewise, as shown in Fig.2(b), the normalized function of the
double slit is
\begin{equation}
\label{eq10} \Psi(x)=\frac{1}{\sqrt{100}}, \ \mbox{ }-75<x<-25,\
\mbox{ } 25<x<75
\end{equation}
\begin{equation}
\label{eq11}\Psi(x)=0, \ \mbox{ }|x|>75,\ \mbox{ } |x|<25
\end{equation}
The normalized momentum distributions of the nanoparticle with
different values of $\alpha$ in the double slit interference are
calculated in the same way as above for the single slit. The
distributions $P(p_x^{(\alpha)})$ are shown in Fig.4.
\begin{figure}[htbp]
\centerline{\includegraphics[width=4.5in,height=2.448in]{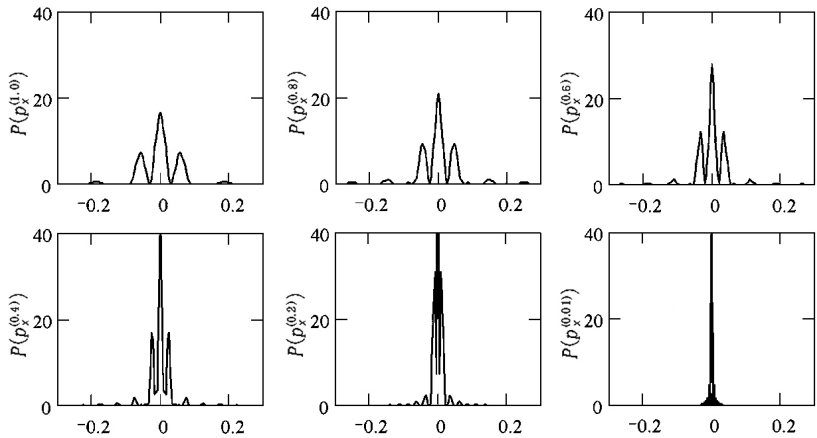}}
\label{Fig.4} \caption{The normalized momentum distribution
$P(p_x^{(\alpha)})$ of the nanoparticle in the double slit
interference with different effective screening mass parameters
$\alpha$.}
\end{figure}
We see the width of the distribution decreases to its classical
limit as $\alpha\rightarrow 0$. Clearly, the diffraction of
nanoparticles from a grating can be calculated in the same way.

The patterns of the single slit diffraction and double slit
interference can be calculated from the momentum distributions if
the momentum ($mv$) of the nanoparticle and the distance between
the slit screen and the detector screen are given.

\section{Nanoparticle inside an infinite square potential well}
Consider an infinite square potential well of width \emph{L}, that
is, the potential function
\begin{equation}
\label{12} V(x)=0, \ \mbox{ } 0<x<L
\end{equation}

\begin{equation}
\label{13} V(x)=\infty, \ \mbox{ } x\leq 0, \ \mbox{ } x\geq L
\end{equation}
As seen in any quantum mechanics textbooks, the quantized energy
and normalized wave function of a particle of mass \emph{m} in the
well are
\begin{equation}
\label{14} E_n=\frac{n^2\pi^2\hbar^2}{2mL^2}, \ \mbox{ }
n=1,2,3,\cdots
\end{equation}

\begin{equation}
\label{15} \psi_n(x)=\sqrt{\frac{2}{L}}\sin(\frac{n\pi x}{L})
\end{equation}
The Fourier transformation of Eq.15 yields the momentum wave
functions
\begin{equation}
\label{16} \phi_n(p)=\frac{1}{\sqrt{2\pi \hbar}}\int
\exp(-\frac{ip x}{\hbar})\psi_n(x)\texttt{d}x
\end{equation}
Now, for a nanoparticle with the ESMP $\alpha$, let \emph{p'}
denote the momentum of the effective screening layer. Substituting
$p'/\alpha$ for \emph{p} into Eq.16 yields
\begin{equation}
\label{17} \Theta_n(p',\alpha)=\frac{1}{\sqrt{2\pi \hbar}}\int
\exp(-\frac{ip' x}{\alpha\hbar})\psi_n(x)\texttt{d}x
\end{equation}
Thus, we have the normalized momentum distribution
\begin{equation}
\label{18} Q_n(p',\alpha)=\frac{|\Theta_n(p',\alpha)|^2}{\int
|\Theta_n(p',\alpha)|^2 \texttt{d}p'}
\end{equation}
The total momentum $p$ of the nanoparticle consists of the quantum
part \emph{p'} and classical part $\pm (1-\alpha)n\pi \hbar /L$,
that is,
\begin{equation}
\label{19} p=p'+(\frac{p'}{|p'|})(1-\alpha)n\pi \hbar /L
\end{equation}
Let $Q^{(\alpha)}_n(p)$ denote the normalized momentum
distribution of the nanoparticle with respect to \emph{p}. It can
be obtained from $Q_n(p',\alpha)$ by making use of the translation
$p'\rightarrow p$. As an illustration example, taking $\hbar=1$ in
the natural unit and $L=1$, the momentum distributions
$Q^{(\alpha)}_2(p)$ are calculated and shown in Fig.5. We see the
distribution approaches to the classical limit $p\rightarrow \pm
2\pi \hbar/L$ as $\alpha \rightarrow 0$.

\begin{figure}[htbp]
\centerline{\includegraphics[width=4.5in,height=2.375in]{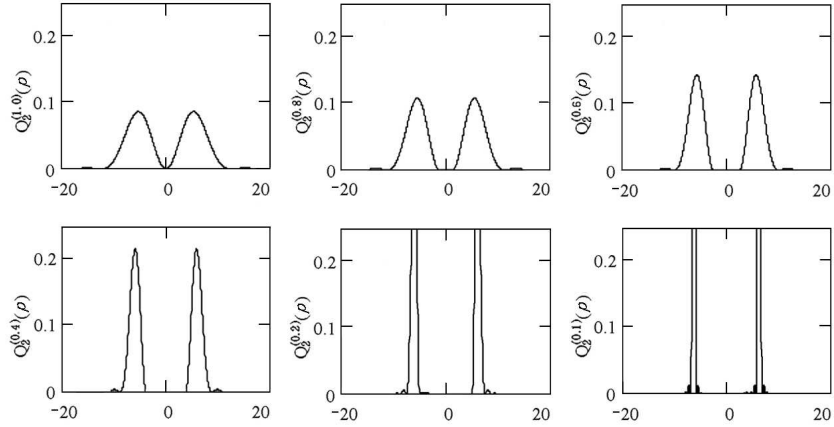}}
\label{Fig.5} \caption{The normalized momentum distributions
$Q_2^{(\alpha)}(p^)$ of the nanoparticle inside the infinite
square potential well with different effective screening mass
parameters $\alpha$.}
\end{figure}

\section{Tunnelling of nanoparticles }
With regard to tunnelling effects, we consider a particle with
momentum $Mv$ and kinetic energy $E=Mv^2/2$ passing through or
over a one-dimensional square potential barrier of height $U_0$
and thickness \emph{L}. According to quantum mechanics, its wave
function can be split into three parts:
\begin{equation}
\label{20} \psi(x)=\exp(ikx)+\sqrt{\rho}\exp(-ikx), \ \mbox{ }
x<0, \ \mbox{ } k=\frac{\sqrt{2ME}}{\hbar}
\end{equation}

\begin{equation}
\label{21} \psi(x)=A\exp(i\kappa x)+B\exp(-i\kappa x),\ \mbox{ }0
\leq x\leq L, \ \mbox{ } \kappa=\frac{\sqrt{2M(E-U_0)}}{\hbar}
\end{equation}

\begin{equation}
\label{22} \psi(x)=\sqrt{\tau}\exp(ik),\ \mbox{ } x>L
\end{equation}
As seen in any quantum mechanics textbooks, for the case where
$E>U_0$, the transmission coefficient is
\begin{equation}
\label{23}
\tau=[1+\frac{U_0^2}{4E(E-U_0)}\sin^2(\frac{\sqrt{2M(E-U_0)}L}{\hbar})]^{-1}
\end{equation}
and for $E<U_0$, substituting $i\kappa$ for $\kappa$ into Eq.21,
the transmission coefficient is
\begin{equation}
\label{24}
\tau=[1+\frac{U_0^2}{4E(U_0-E)}\sinh^2(\frac{\sqrt{2M(U_0-E)}L}{\hbar})]^{-1}
\end{equation}
The reflection coefficient is
\begin{equation}
\label{25} \rho=1-\tau
\end{equation}

Now we are going to calculate the transmission coefficient and
reflection coefficients of a nanoparticle with the ESMP $\alpha$.
Since the momentum of the effective screening layer is
$k^{(\alpha)}\hbar=\alpha k\hbar$ outside the barrier and
$\kappa^{(\alpha)} \hbar=\alpha \kappa\hbar$ inside the barrier,
we have to substitute $k^{(\alpha)}/\alpha$ for $k$ and
$\kappa^{(\alpha)}/\alpha$ for $\kappa$ into the wave functions.
Clearly the substitution is equivalent to substituting $\alpha
\hbar$ for $\hbar$. So, for the case where $E>U_0$, we obtain the
transmission coefficient
\begin{equation}
\label{26} \tau_\alpha=[1+\frac{U_0^2}{4E(E-U_0)}
\sin^2(\frac{\sqrt{2M(E-U_0)}L}{\alpha\hbar})]^{-1}
\end{equation}
and for $E<U_0$, similarly, we have
\begin{equation}
\label{27} \tau_\alpha=[1+\frac{U_0^2}{4E(U_0-E)}
\sinh^2(\frac{\sqrt{2M(U_0-E)}L}{\alpha\hbar})]^{-1}
\end{equation}
The reflection coefficient is
\begin{equation}
\label{28} \rho_\alpha=1-\tau_\alpha
\end{equation}
For example, for the case where $E=1$ and $U_0=2$, taking $L=1$
and $M=0.1, 0.5, 1.0$, the transmission coefficient curves
$\tau(\alpha,M)$ are shown in Fig.6(a). If taking $M=1.0$ and
$L=0.1, 0.5, 1.0$, the curves $\tau(\alpha,L)$ are shown in
Fig.6(b). Here we have taken $\hbar=1$ in the natural unit.
\begin{figure}[htbp]
\centerline{\includegraphics[width=3.6in,height=1.765in]{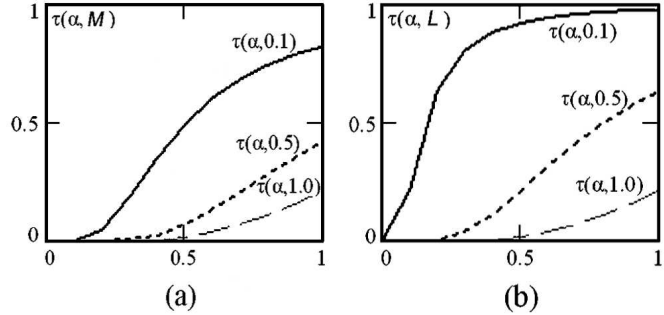}}
\label{Fig.6} \caption{Transmission coefficients of the
nanoparticle dependent on $\alpha$ for the case where $E<U_0$: (a)
the curves $\tau(\alpha,M)$ with different values of \emph{M}, (b)
the curves $\tau(\alpha,L)$ with different values of \emph{L}.}
\end{figure}
We see the transmission coefficients approach to their classical
limits as $\alpha\rightarrow 0$.

\section{Uncertainty relations for nanoparticles}
In 1927, Heisenberg stated: $``$the more precisely the position is
determined, the less precisely the momentum is known in this
instant, and vice vera.$"$ [6] This rule is known as Heisenberg
uncertainty principle. Its physical reasoning has been offered in
Ref.[2]. The principle is basically formulated by the relation
\begin{equation}
\label{eq29} \triangle p_x\triangle x\geq\frac{\hbar}{2}
\end{equation}
Now, for a nanoparticle with the ESMP $\alpha$, since $\triangle
p_x^{(\alpha)}=\alpha \triangle p_x$ as seen from Fig.1(b), the
Heisenberg relation becomes
\begin{equation}
\label{eq30} \triangle p_x^{(\alpha)}\triangle x=\triangle
p_x\alpha\triangle x\geq\frac{\alpha\hbar}{2}
\end{equation}
Figs.3-6 illustrate the momentum uncertainty of nanoparticles,
which decreases with decreasing $\alpha$. We would have similar
energy-time, angle momentum-angle and other uncertainty relations.
Thus the Heisenberg uncertainty principle is in some degree
dependent on $\alpha$ applicable to nanoparticles and completely
unapplicable to macroscopic objects. The fact that the momentum
and position are exactly measurable in classical physics reflects
the limit $\alpha \rightarrow0$, instead of wrong limit $\hbar
\rightarrow 0$.  It is a logical mistake to regard $\hbar$ as a
variable instead of a definite constant.

\section{Conclusion}
We have calculated the momentum distributions of nanoparticles in
diffraction and interference dependent on the effective screening
mass parameter or size parameter and presented the calculations
for a nanoparticle inside an infinite square potential well and
for a tunnelling nanoparticle through a square potential barrier.
These results display the transition from quantum to classical
mechanics and the simultaneous wave-particle duality of
nanoparticles. The concept that the effective screening effect
increases with increasing size of an object paves way for
development of nanomechanics and nanotechnology.

\end{document}